\documentclass{elsart3}

\usepackage{graphicx}

\begin{document}

\begin{frontmatter}

\title{Meissner effect in diffusive normal metal / $d$-wave
superconductor junctions}

\author[address1]{Takehito Yokoyama\thanksref{thank1}},
\author[address1]{Yukio Tanaka},
\author[address2]{Alexander Golubov}
\author[address1]{Jun-ichiro Inoue},
\author[address3]{Yasuhiro Asano},

\address[address1]{Department of Applied Physics, Nagoya University, Nagoya 464-8603, Japan.\\ CREST Japan Science and Technology Corporation (JST), Nagoya 464-8603, Japan.}

\address[address2]{ Faculty of Science and Technology, University of Twente, 7500 AE,
Enschede, The Netherlands.}

\address[address3]{ Department of Applied Physics, Hokkaido University, Sapporo 060-8628, Japan.}

\thanks[thank1]{Corresponding author. Present address: Department of Applied
Physics,  Nagoya University, Nagoya, Japan\\
 E-mail: h042224m@mbox.nagoya-u.ac.jp}

\begin{abstract}
The Meissner effect in diffusive normal metal / insulator /  $d$-wave
superconductor junctions is studied theoretically in the framework
of the Usadel equation under the generalized boundary condition. The
effect of midgap Andreev resonant states (MARS) formed at the
interface of $d$-wave superconductor is taken into account. It is
shown that the formation of MARS suppresses the susceptibility of
the diffusive normal metal.
\end{abstract}
\begin{keyword}
Meissner effect; proximity effect; midgap Andreev resonant states; $d$-wave superconductor
\end{keyword}
\end{frontmatter}

\section{Introduction}
In diffusive normal metal / superconductor (DN/S) junctions, the DN
acquires induced superconductivity, i.e. Cooper pairs penetrate into
the DN. This proximity effect has been studied since the BCS theory
was established. The proximity induced Meissner demagnetization in
DN/S junctions was measured experimentally by Oda and
Nagano\cite{Oda} and Mota et al.\cite{Mota1}. It has $T^{-1/2}$
dependence in the dirty limit. The quasiclassical Green's function
theory was used earlier to study the Meissner effect in proximity
structures.

The quasiclassical Green's function theory was developed by 
Eilenberger~\cite{Eilenberger} and was generalized by Eliashberg~\cite{Eliashberg}, 
Larkin and Ovchinnikov~\cite {Larkin} in order to study the nonequilibrium
state. This theory was applied by Zaikin\cite{Zaikin} and
Kieselmann\cite{Kieselmann} to studing the Meissner effect in DN/S
junctions. Narikiyo and Fukuyama~\cite{Narikiyo} calculated the
Meissner screening length in a semi-infinite system containing
an Anderson impurity. Higashitani and Nagai studied the Meissner effect
in the clean limit~\cite{Higashitani}. Belzig et al.~\cite{Bel1,Bel2}
have considered more realistic systems by assuming a perfectly transparent
N/S interface. Up to now the boundary conditions derived by
Kupriyanov and Lukichev (KL) \cite{KL} were widely used to study
proximity effect in DN/S structures.

%
%
A more general boundary conditions was derived by Nazarov
\cite{Nazarov2} based on the  Keldysh-Nambu Green's function
formalism \cite{Zaitsev} within the framework of the
Landauer-B\"{u}ttiker scattering formalism.
The merit of this boundary condition is that the BTK
theory\cite{BTK} is reproduced in the ballistic limit while in the
diffusive limit with a low transmissivity of the interface, the KL
boundary condition is reproduced.
Although almost all previous papers on Meissner effect in
mesoscopic NS junctions are either based on the KL boundary
conditions or on the BTK model, in the actual junctions,
the transparency of the junction is not always small and 
the impurity scattering in the DN cannot be neglected. 
Tanaka et al.\cite{TGK} and
Yokoyama et al.\cite{Yoko} calculated tunneling conductance by using
the Nazarov's boundary condition.

It is well
known in $d$-wave superconductors that the midgap Andreev resonant
states (MARS) are formed at the interface of $d$-wave
superconductor. The MARS crucially influence various physical quantities
\cite{TK}. One of the authors (Y.T.) recently
generalized the boundary condition of the
Keldysh-Nambu Green's function formalism to unconventional
superconductor junctions~\cite{TNGK,pwave}.
It is revealed that in DN/$d$-wave superconductor junctions 
the proximity effect and the MARS strongly compete with each other~\cite{TNGK},
while they coexist in DN/triplet superconductor junctions.
The newly obtained boundary conditions expressed in the Keldysh-Nambu
Green's function are useful for the calculation of various physical
quantities.
The timely problem is to study theoretically the Meissner effect in
DN / $d$-wave S junctions using the new boundary conditions
\cite{TNGK}. In the present paper, we calculate the susceptibility
of the DN layer in DN/ $d$-wave S junctions for various junction parameters
such as the height of the insulating barrier at the interface and
the angle between the normal to the interface and the crystal axis
of a $d$-wave superconductor.
\par
The organization of the paper is as follows. In section 2, we will
provide the derivation of the expression for the susceptibility of
the DN. In section 3, the results of calculation are presented for
various types of junction. In section 4, the summary of the obtained
results is given. In the present paper we set $c=k_B=\hbar=1$.

\section{Formulation}

In this section, we introduce the model and the formalism. We
consider a junction consisting of vacuum (VAC) and superconducting
reservoirs connected by a quasi-one-dimensional diffusive conductor
(DN) with a length $L$ much larger than the mean free path. We
assume that the interface between the DN conductor and the S
electrode at $x=L$ has a resistance $R_{b}$, the DN/VAC interface at
$x=0$ is specular, and we apply the generalized boundary conditions
by Tanaka \cite{TNGK} to treat the interface between DN and S. A
weak external magnetic field $H$ is applied in $z$-direction (see
Fig. 1). The vector potential can be chosen to have only the $y$
component which depends on $x$.
\begin{figure}[htb]
\begin{center}
\scalebox{0.4}{
\includegraphics[width=18.0cm,clip]{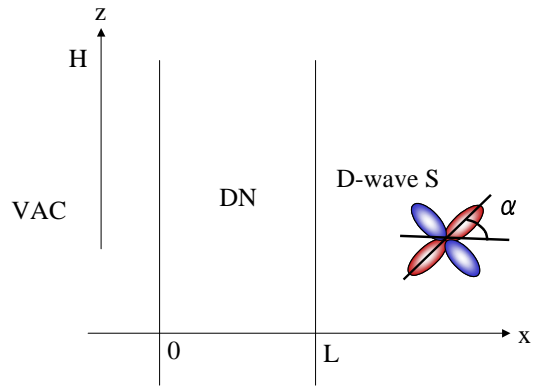}}
\end{center}
\par
\caption{Schematic illustration of the model.}
\end{figure}

We describe the insulating barrier between DN and S by using the $\delta$-function 
(i.e., $U(x)=H\delta(x-L)$), which provides the transparency of the
junction $T_{m}=4\cos ^{2}\phi /(4\cos ^{2}\phi +Z^{2})$, where
$Z=2H/v_{F}$ is a dimensionless constant, $\phi $ is the
injection angle of a quasiparticle measured from the interface normal to the junction
and $v_{F}$ is Fermi velocity.

In the following, we solve the Usadel equations
\cite{Usadel}
with using the standard $\theta$-parameterization. 
The parameter $\theta (x)$ is a measure of the proximity effect in DN and 
obey the following equation
\begin{equation}
D\frac{\partial ^{2}}{\partial x^{2}}\theta (x)-2\omega_n\sin [\theta (x)]=0,  \label{Usa1}
\end{equation}
where $D$ and $\omega_n$ denote the diffusion constant and the Matsubara frequency, respectively.
The boundary condition for $\theta(x)$ at the DN/S interface is given in Ref.~\cite{TNGK}. The
interface resistance $R_{b}$ is given by
\begin{equation}
R_{b}=R_{0} \frac{2} {\int_{-\pi/2}^{\pi/2} d\phi T(\phi)\cos\phi}
\end{equation}
with $ T(\phi)=4\cos ^{2}\phi /(4\cos ^{2}\phi +Z^{2})$.
Here $R_{0}$ is Sharvin resistance $R_{0}^{-1}=e^{2}k_{F}^2S_c/(4\pi^{2})$, where $k_{F}$ is the 
Fermi wave number and $S_c$ is the constriction area.
The current distribution is given by
\begin{equation}
j(x) =  - 8\pi e^2 N\left( 0 \right)DT\sum\limits_{\omega _n  > 0}
{\sin ^2 \theta \left( x \right)} A\left( x \right),
\end{equation}
where $A(x)$, $N(0)$ and $T$ denote the vector potential, the density of states at the Fermi 
energy and the temperature of the system respectively.
The Maxwell equation reads
\begin{equation}
\frac{{d^2 }}{{dx^2 }}A\left( x \right) =  - 4\pi j\left( x \right).
\end{equation}
The boundary conditions for $A(x)$ are given by
\begin{eqnarray}
 \frac{d}{{dx}}A\left( 0 \right) = H, \qquad  A\left( L \right) = 0,
\end{eqnarray}
where we have neglected the penetration of magnetic fields into the
superconductor by assuming a small penetration depth in S.

Finally we obtain the expression of the susceptibility,
\begin{equation}
- 4\pi \chi  = 1 + \frac{{A\left( 0 \right)}}{{HL}}.
\end{equation}
The $d$-wave pair potentials in directional space are given by
$\Delta_{\pm} = \Delta(T)\cos2(\phi \mp \alpha)$, where $\Delta(T)$
is the magnitude of pair potential at a given temperature $T$ and
$\alpha$ denotes the angle between the normal to the interface and
the crystal axis of a $d$-wave superconductor.

\section{Results}
In the following, we focus on the magnitude of the diamagnetic
susceptibility $\chi$ induced by the proximity effect.
Figs. 2 and 3 show the susceptibility for $Z=10$ and $Z=0$ 
respectively  where $K =16\pi e^2 N\left( 0 \right)D^2$.
For $\alpha=0$, the temperature dependencies of
$-4\pi\chi$ are not much different.
For $\alpha=0.125\pi$, the magnitude of $\chi$ for $Z=10$ is much
 stronger suppressed than that for $Z=0$. 
At the same time, we find that the magnitude of $\chi$ decreases
with increasing $\alpha$.
We note in the case of $\alpha=0.25\pi$ that the susceptibility
completely vanishes, (i.e., $-4\pi\chi=0$).
This is because the proximity effect is absent in diffusive metals due to angular averaging\cite{TNGK}.
The absence of the proximity effect is a significant
feature specific for junctions containing unconventional superconductors.

\begin{figure}[htb]
\begin{center}
\scalebox{0.4}{
\includegraphics[width=16.0cm,clip]{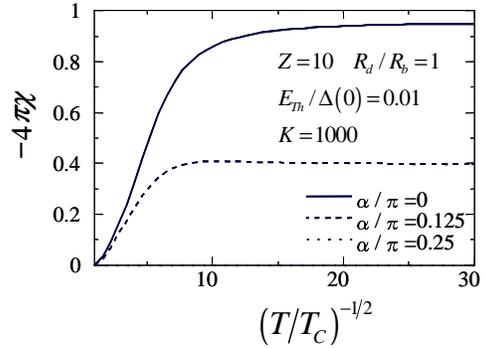}}
\end{center}
\par
\caption{Susceptibility for low transparent junctions with $Z=10$.}
\end{figure}

\begin{figure}[htb]
\begin{center}
\scalebox{0.4}{
\includegraphics[width=16.0cm,clip]{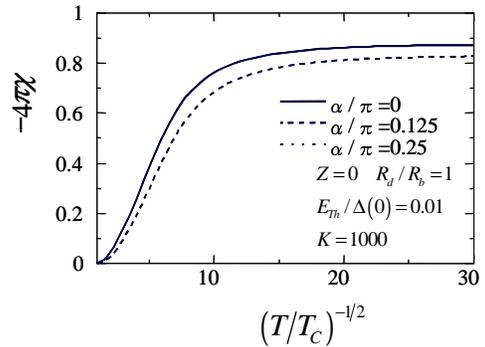}}
\end{center}
\par
\caption{ Susceptibility for high transparent junctions with $Z=0$.}
\end{figure}

 We also plot the $\alpha$ dependencies of the susceptibility at 
$T/T_C=0.01$ and $T/T_C=0.1$ in Fig. 4.
For all cases, $\chi$ is a decreasing function of $\alpha$.
At $T/T_C=0.01$, the magnitude of $\chi$ for $Z=10$ rapidly
decreases with the increase of $\alpha$.
The results imply that the MARS  suppresses the proximity effect 
in low transparent junctions and low temperatures.

\begin{figure}[htb]
\begin{center}
\scalebox{0.4}{
\includegraphics[width=16.0cm,clip]{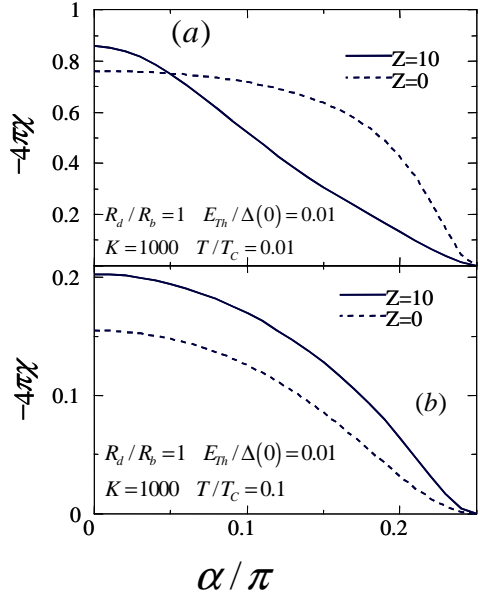}}
\end{center}
\par
\caption{ $\alpha$ dependences of the susceptibility at $T/T_C=0.01$(upper panel) and $T/T_C=0.1$(lower panel).}
\end{figure}


\section{Conclusions}
In the present paper, we have calculated the induced
Meissner effect by the proximity effect in
DN region of DN/$d$-wave superconductor junctions.
We have solved the Usadel equation under a general boundary condition \cite{TNGK}
in which the formation of the MARS is fully taken into account~\cite{TK}.
The magnitude of $\chi$ decreases with the increase of
$\alpha$ up to $0.25\pi$.
At $\alpha=0.25\pi$, where all quasiparticles feel MARS, the
$\chi$ becomes zero.
It might be interesting to check experimentally such an anomalous
proximity effect in DN.
Another future problem is a similar calculation of the induced
Meissner effect with a $p$-wave triplet superconductor 
instead of a $d$-wave one, since dramatic new phenomena are recently
predicted in DN/ triplet junctions \cite{pwave}.

The authors appreciate useful and fruitful discussions with Yu. Nazarov and
H. Itoh. This work was supported by the Core Research for Evolutional
Science and Technology (CREST) of the Japan Science and Technology
Corporation (JST). The computational aspect of this work has been performed
at the facilities of the Supercomputer Center, Institute for Solid State
Physics, University of Tokyo and the Computer Center.
%


\end{document}